\documentclass[fp,twocolumn]{jpsj3}
\usepackage{txfonts}

\usepackage{color}

\title{Root mean squares of distance and geodesic between two constituent particles within fractal aggregates prepared by BCCA, DLA, and GSAW procedures}

\author{Sota Arakawa\thanks{arakawas@jamstec.go.jp}}
\inst{Yokohama Institute for Earth Sciences, Japan Agency for Marine-Earth Science and Technology, \\
 3173-25, Showa-machi, Kanazawa-ku, Yokohama, Kanagawa, 236-0001, Japan} 

\abst{
Understanding the geodesic properties of fractal aggregates is essential, as their thermal and mechanical properties are characterized by their geodesics.
In this study, we investigate the root mean square (RMS) of the geodesic between two constituent particles within fractal aggregates prepared by ballistic cluster--cluster aggregation (BCCA), diffusion-limited aggregation (DLA), and growing self-avoiding walk (GSAW) processes in two- and three-dimensional spaces.
We find that the dependence of the RMS of the geodesic on the number of constituent particles is given by the following equation: $N \approx k_{\rm g} {D_{\rm RMS}}^{d_{\rm g}}$, where $N$ is the number of constituent particles and $D_{\rm RMS}$ is the RMS of the geodesic.
We numerically obtain the prefactor $k_{\rm g}$ and exponent $d_{\rm g}$ for these fractal aggregates.
We name the exponent ``the geodesic dimension'', and it is compared with the fractal dimension.
Our findings show that the difference between fractal and geodesic dimensions varies significantly depending on the preparation procedure for fractals.
}

\begin{document}
\maketitle

\section{Introduction}

The agglomeration of small particles to grow into larger aggregates is a key process in many fields of science and engineering \cite{enustun1963coagulation, 1985prpl.conf.1100H, frenklach1991detailed}.
The resulting aggregates frequently have a complex and porous structure.
Moreover, porous aggregates formed via some procedures are fractals.

Fractals are ubiquitous in nature \cite{1967Sci...156..636M, 1982fgn..book.....M}.
The geometrical structure of fractal aggregates depends on their coagulation process \cite{meakin1999historical}.
Impacts of the geometrical structure on the physical properties of aggregates have been intensively investigated.
For example, the optical properties of fractal aggregates strongly depend on their structure \cite{2021MNRAS.504.2811T}.
The fractal structure also controls the thermal and mechanical properties of fractal aggregates \cite{2013A&A...554A...4K, 2019PTEP.2019i3E02A, 2019ApJ...874..159T}.

The structure of a fractal aggregate is frequently characterized by the fractal dimension.
The fractal dimension is the exponent of the relationship between the radius of aggregates and the number of constituent particles.
The fractal dimension defined by the root mean square (RMS) of the distance between two constituent particles, $d_{\rm f}$, is not always equal to the spatial dimension, and the density of aggregates decreases with increasing number of constituent particles when the fractal dimension is smaller than the spatial dimension.
The fractal dimensions of various aggregate structures have been reported in the literature \cite{meakin1999historical, 1992A&A...262..315M, 2009JPSJ...78a3605O, 2009ApJ...707.1247O}.

The geodesic within a fractal aggregate is another important property.
Indeed, the thermal and mechanical properties of fractal aggregates are characterized by the geodesic within the aggregates \cite{2019PTEP.2019i3E02A}.
However, the geodesic properties of fractal aggregates have been poorly understood in comparison with their radius or fractal dimension.
For aggregates formed via the ballistic cluster--cluster aggregation (BCCA) process in the three-dimensional space, the exponent of the relationship between the RMS of the geodesic and the number of constituent particles (hereinafter referred to as the geodesic dimension, $d_{\rm g}$) is $d_{\rm g} \approx 1.4$, whereas the fractal dimension is $d_{\rm f} \approx 1.9$ \cite{2019PTEP.2019i3E02A}.
The difference between two exponents might be a common feature among various fractal aggregates, although to the best of our knowledge, it has never been investigated.

In this study, we numerically investigate the fractal and geodesic dimensions of aggregates prepared by various procedures in two- and three-dimensional spaces.  
As analogues of natural fractals, we consider three types of probabilistic coagulation process: BCCA, diffusion-limited aggregation (DLA), and growing self-avoiding walk (GSAW) processes.
We also study a mathematical fractal called the Vicsek fractal formed by a deterministic process as a reference.

The remainder of this paper is organized as follows.
In Section \ref{sec:model}, we introduce the distance and geodesic within an aggregate and define the fractal and geodesic dimensions.
In Sections \ref{sec:BCCA}--\ref{sec:Vicsek}, we present the results of numerical simulations for each preparation procedure.
Finally, Section \ref{sec:summary} is dedicated to summarizing and discussing the results.

\section{Fractal and Geodesic Dimensions}
\label{sec:model}

For the quantitative analysis of cluster structures, it is essential to define the sizes of clusters.
In this study, we calculate the sizes of clusters using the RMSs of the distance and geodesic between two constituent particles.

\begin{figure}
\centering
\includegraphics[width = 0.6\columnwidth]{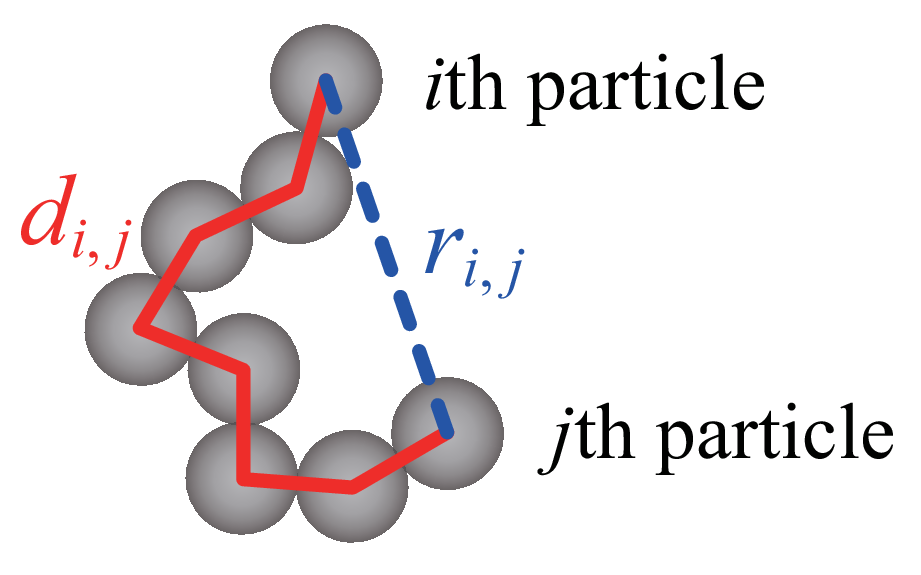}
\caption{\label{fig:schematic}
Schematic of the distance between the $i$th and $j$th particles, $r_{i, j}$, and the geodesic between the $i$th and $j$th particles, $d_{i, j}$.
The radius of each particle is $1$.
(Color online)
}
\end{figure}

For this study, the radius of each particle is set to be $1$.
Figure \ref{fig:schematic} illustrates the distance and geodesic between the $i$th and $j$th particles in a cluster denoted as $r_{i, j}$ and $d_{i, j}$, respectively.
When two particles are in contact, both the distance and geodesic are $2$.
It is evident that the relationship $r_{i, j} \le d_{i, j}$ is always satisfied by definition.

The RMS of the distance between two constituent particles, $R_{\rm RMS}$, is given by
\begin{equation}
R_{\rm RMS} = {\left( \frac{1}{N^{2}} \sum_{i = 1}^{N} \sum_{j = 1}^{N} {r_{i, j}}^{2} \right)}^{1/2},
\end{equation}
where $N$ is the number of constituent particles.
We note that $R_{\rm RMS}$ is closely related to the gyration radius and it is given by $R_{\rm gyr} = R_{\rm RMS} / \sqrt{2}$.
Similarly, the RMS of the geodesic is given by
\begin{equation}
D_{\rm RMS} = {\left( \frac{1}{N^{2}} \sum_{i = 1}^{N} \sum_{j = 1}^{N} {d_{i, j}}^{2} \right)}^{1/2}.
\end{equation}
It is clear that the relation $D_{\rm RMS} \ge R_{\rm RMS}$ is satisfied for arbitrary clusters.

When aggregates are fractal, the ensemble average of $\ln{R_{\rm RMS}}$ should be given by the following equation for a sufficiently large $N$:
\begin{equation}
{\langle \ln{R_{\rm RMS}} \rangle} = \frac{1}{d_{\rm f}} \ln{\left( \frac{N}{k_{\rm f}} \right)},
\end{equation}
where $d_{\rm f}$ is the fractal dimension and $k_{\rm f}$ is the fractal prefactor.
In other words, the relationship between $N$ and $R_{\rm RMS}$ is approximately given by $N \approx k_{\rm f} {R_{\rm RMS}}^{d_{\rm f}}$.
Similarly, the ensemble average of $\ln{D_{\rm RMS}}$ is given by
\begin{equation}
{\langle \ln{D_{\rm RMS}} \rangle} = \frac{1}{d_{\rm g}} \ln{\left( \frac{N}{k_{\rm g}} \right)}.
\end{equation}
Here, we call $d_{\rm g}$ the geodesic dimension and $k_{\rm g}$ the geodesic prefactor.
As $D_{\rm RMS} \ge R_{\rm RMS}$ is satisfied for an arbitrary $N$, the following relation would hold true for any fractal aggregates:
\begin{equation}
d_{\rm g} \le d_{\rm f}.
\end{equation}

We note that the geodesic within a cluster is essentially identical to the path length within a graph network.
The average and frequency distribution of path lengths of complex networks have been studied in various fields of science, including those on cellular networks, the Internet, and networks in linguistics \cite{2002RvMP...74...47A, 2002PNAS...9913382Y}.

\section{Ballistic Cluster--Cluster Aggregation}
\label{sec:BCCA}

\begin{figure}
\centering
\includegraphics[width = \columnwidth]{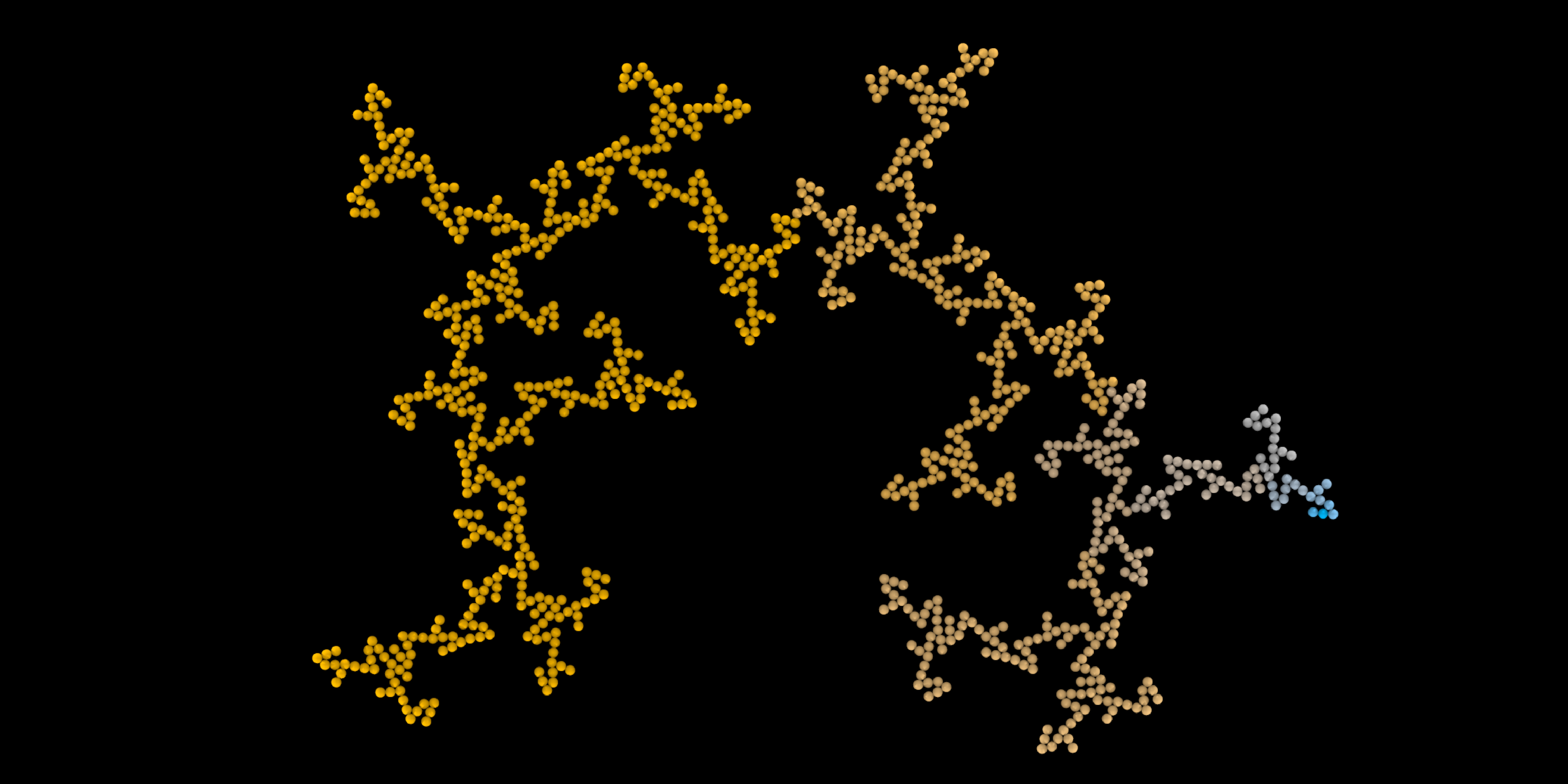}
\caption{\label{fig:BCCA_2d}
Snapshot of a cluster formed by BCCA in two-dimensional space.
The color of each particle represents the timing of deposition; blue particles are deposited on the cluster at an earlier stage than yellow ones.
The number of constituent particles is $N = 1024$.
(Color online)
}
\end{figure}

BCCA is a sequence of successive collisions between two identical clusters \cite{1971ChEnS..26.2071S, 1984JPhA...17L.771J, 1988JChPh..89..246M}.
At each growth step, we randomly choose the impact parameter and the orientations of the clusters.
Then, we calculate the ballistic trajectory until one of the constituent particles collides with another particle, and we copy the newly formed cluster.
Figure \ref{fig:BCCA_2d} shows a snapshot of a cluster formed by BCCA in the two-dimensional space.
The number of constituent particles in a cluster formed after $i$ steps is $N = 2^{i}$.

The geometrical properties of clusters formed by BCCA are intensively investigated in the fields of astronomy and astrophysics \cite{1992A&A...262..315M, 2009ApJ...707.1247O, 2019ApJ...874..159T, 2021MNRAS.504.2811T}.
In the early stage of protoplanetary dust growth, large and porous dust aggregates are formed by low-speed sticking collisions between two similarly sized clusters \cite{2000PhRvL..85.2426B}.
Therefore, the initial dust aggregates in protoplanetary disks, which are the building blocks of planets, may resemble clusters formed by BCCA.

In this study, we perform 50 runs of sequential collisions of clusters for both two- and three-dimensional cases.
The final size of clusters is $N = 2^{20} = 1048576$, and we calculate the exponents ($d_{\rm f}$ and $d_{\rm g}$) and prefactors ($k_{\rm f}$ and $k_{\rm g}$) using the data for $2^{12} \le N \le 2^{20}$.

\begin{figure}
\centering
\includegraphics[width = \columnwidth]{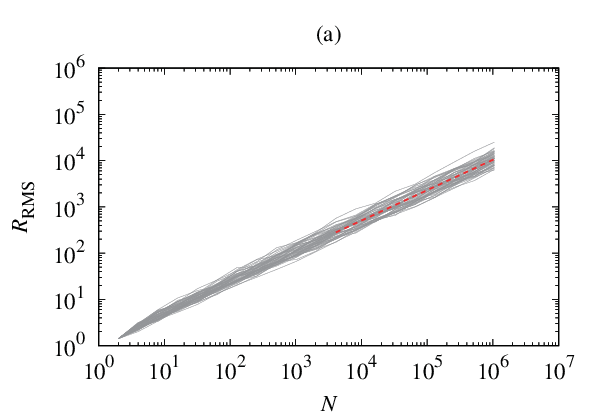}
\includegraphics[width = \columnwidth]{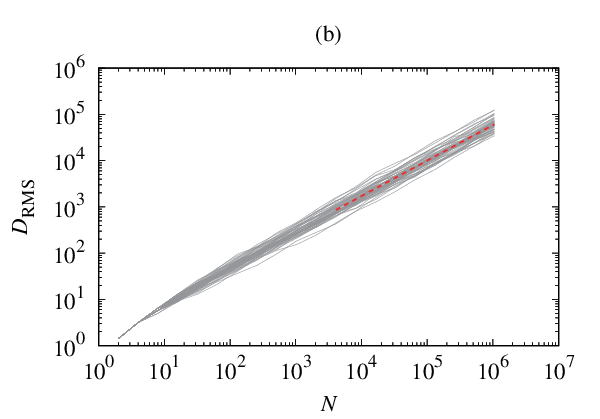}
\caption{\label{fig:N_R_D_BCCA_2d}
Dependences of (a) $R_{\rm RMS}$ and (b) $D_{\rm RMS}$ on the number of constituent particles within aggregates formed by BCCA in two-dimensional space.
Gray solid lines show the evolutionary paths of 50 aggregates, and the red dashed line represents the best fit obtained by the least-squares method.
(Color online)
}
\end{figure}

Figure \ref{fig:N_R_D_BCCA_2d} shows the dependences of $R_{\rm RMS}$ and $D_{\rm RMS}$ on the number of constituent particles within aggregates formed by BCCA in the two-dimensional space.
Gray solid lines show the evolutionary paths of 50 aggregates, and the red dashed line represents the best fit obtained by the least-squares method.
The best fits are ${\langle \ln{R_{\rm RMS}} \rangle} = 0.6570(75) \ln{N} + 0.174(84)$ and ${\langle \ln{D_{\rm RMS}} \rangle} = 0.7676(76) \ln{N} + 0.373(85)$, and the given uncertainties represent standard errors.
We obtain the following results:
\begin{equation}
{\langle \ln{R_{\rm RMS}} \rangle} = \frac{1}{1.52} \ln{\left( \frac{N}{0.767} \right)},
\end{equation}
and
\begin{equation}
{\langle \ln{D_{\rm RMS}} \rangle} = \frac{1}{1.30} \ln{\left( \frac{N}{0.615} \right)}.
\end{equation}
The fractal dimension reported in previous studies \cite{1984JPhA...17L.771J, meakin1999historical} is $d_{\rm f} \approx 1.51$, which is consistent with our result.
To the best of our knowledge, the geodesic dimension $d_{\rm g} = 1.30$ has not been reported; in this study, we found that $d_{\rm g}$ is clearly smaller than $d_{\rm f}$ for two-dimensional BCCA clusters.

\begin{figure}
\centering
\includegraphics[width = \columnwidth]{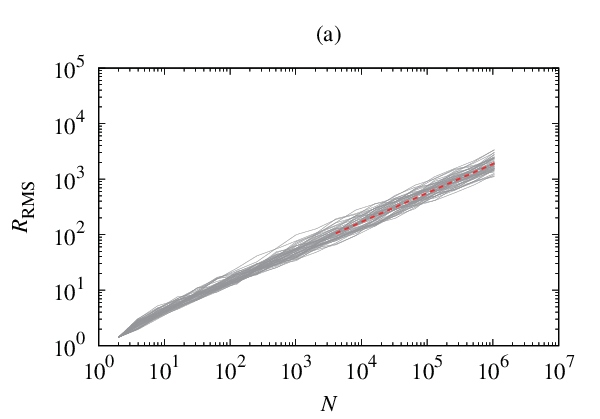}
\includegraphics[width = \columnwidth]{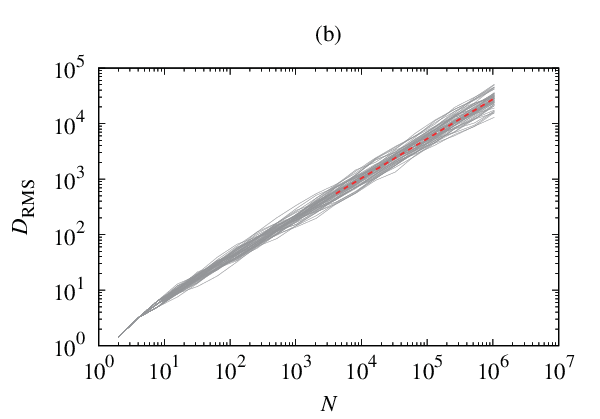}
\caption{\label{fig:N_R_D_BCCA_3d}
Dependences of (a) $R_{\rm RMS}$ and (b) $D_{\rm RMS}$ on the number of constituent particles within aggregates formed by BCCA in three-dimensional space.
Gray solid lines show the evolutionary paths of 50 aggregates, and the red dashed line represents the best fit obtained by the least-squares method.
(Color online)
}
\end{figure}

Figure \ref{fig:N_R_D_BCCA_3d} shows the dependences of $R_{\rm RMS}$ and $D_{\rm RMS}$ on the number of constituent particles within aggregates formed by BCCA in the three-dimensional space.
Gray solid lines show the evolutionary paths of 50 aggregates, and the red dashed line represents the best fit obtained by the least-squares method.
The best fits are ${\langle \ln{R_{\rm RMS}} \rangle} = 0.5224(64) \ln{N} + 0.316(72)$ and ${\langle \ln{D_{\rm RMS}} \rangle} = 0.7118(67) \ln{N} + 0.387(75)$, and the given uncertainties represent standard errors.
We obtain the following results:
\begin{equation}
{\langle \ln{R_{\rm RMS}} \rangle} = \frac{1}{1.91} \ln{\left( \frac{N}{0.546} \right)},
\end{equation}
and
\begin{equation}
{\langle \ln{D_{\rm RMS}} \rangle} = \frac{1}{1.40} \ln{\left( \frac{N}{0.581} \right)}.
\end{equation}
The fractal dimension reported in previous studies \cite{1988JChPh..89..246M, 2009ApJ...707.1247O, 2019PTEP.2019i3E02A} is $d_{\rm f} \approx 1.9$, which is consistent with our result.
The geodesic dimension $d_{\rm g} = 1.40$ is also consistent with that obtained in our previous study \cite{2019PTEP.2019i3E02A}, and we confirm that $d_{\rm g}$ is clearly smaller than $d_{\rm f}$ for both two- and three-dimensional BCCA clusters.

\section{Diffusion-Limited Aggregation}
\label{sec:DLA}

\begin{figure}
\centering
\includegraphics[width = \columnwidth]{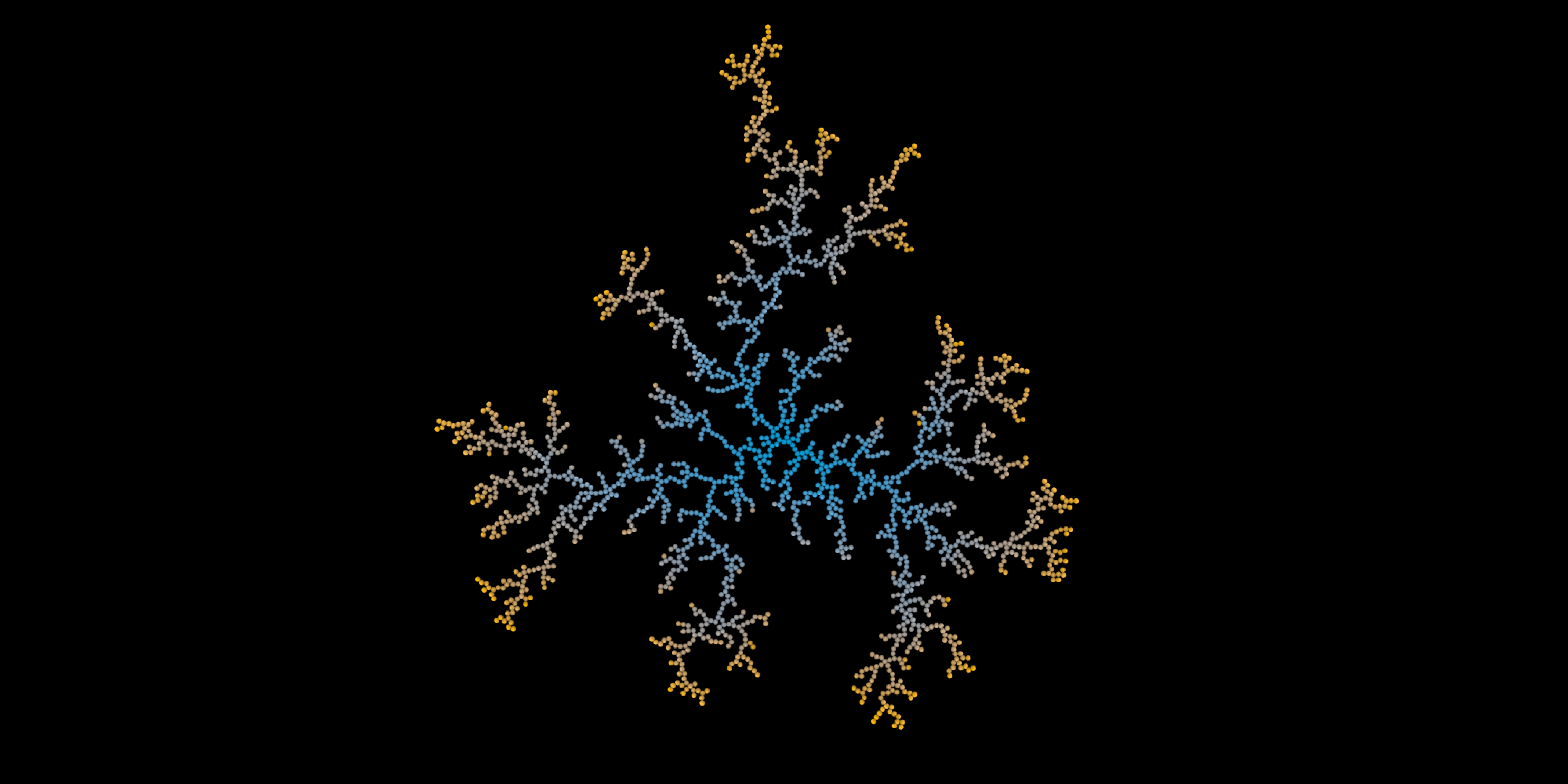}
\caption{\label{fig:DLA_2d}
Snapshot of a cluster formed by DLA in two-dimensional space.
The color of each particle represents the timing of deposition; blue particles are deposited on the cluster at an earlier stage than yellow ones.
The number of constituent particles is $N = 2048$.
(Color online)
}
\end{figure}

DLA is a sequence of successive collisions between a particle and a cluster \cite{1981PhRvL..47.1400W}.
In the DLA procedure, a particle collides with a growing cluster via a random walk path that starts far from the origin.
In this study, we set that the length of each step is $1$, which is equal to the particle radius.
Figure \ref{fig:DLA_2d} shows a snapshot of a cluster formed by DLA in the two-dimensional space.
As cluster growth in a diffusion field is a common phenomenon in nature, the fractal nature of DLA clusters has been intensively investigated \cite{2005PhRvE..72a1406P, 2009JPSJ...78a3605O}.

In this study, we perform 50 runs of sequential collisions of clusters for both two- and three-dimensional cases.
The final size of clusters is $N = 2^{18} = 262144$, and we calculate the exponents ($d_{\rm f}$ and $d_{\rm g}$) and prefactors ($k_{\rm f}$ and $k_{\rm g}$) using the data for $2^{12} \le N \le 2^{18}$.

\begin{figure}
\centering
\includegraphics[width = \columnwidth]{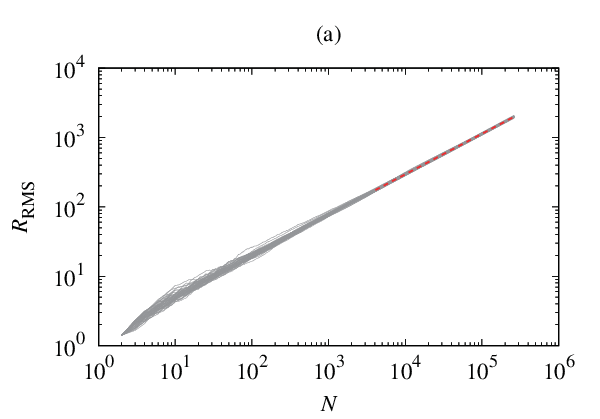}
\includegraphics[width = \columnwidth]{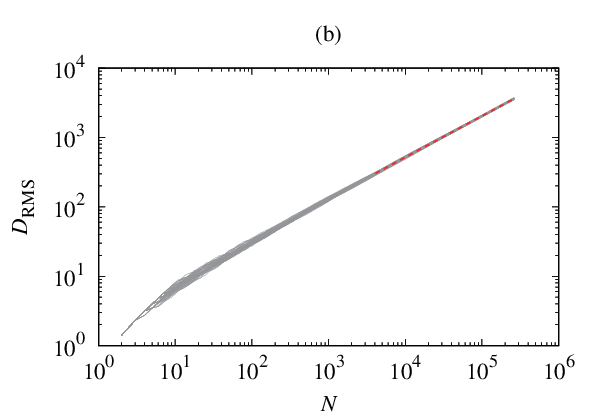}
\caption{\label{fig:N_R_D_DLA_2d}
Dependences of (a) $R_{\rm RMS}$ and (b) $D_{\rm RMS}$ on the number of constituent particles within aggregates formed by DLA in two-dimensional space.
Gray solid lines show the evolutionary paths of 50 aggregates, and the red dashed line represents the best fit obtained by the least-squares method.
(Color online)
}
\end{figure}

Figure \ref{fig:N_R_D_DLA_2d} shows the dependences of $R_{\rm RMS}$ and $D_{\rm RMS}$ on the number of constituent particles within aggregates formed by DLA in the two-dimensional space.
Gray solid lines show the evolutionary paths of 50 aggregates, and the red dashed line represents the best fit obtained by the least-squares method.
The best fits are ${\langle \ln{R_{\rm RMS}} \rangle} = 0.5838399(67) \ln{N} + 0.308474(78)$ and ${\langle \ln{D_{\rm RMS}} \rangle} = 0.5951207(55) \ln{N} + 0.767299(64)$, and the given uncertainties represent standard errors.
We obtain the following results:
\begin{equation}
{\langle \ln{R_{\rm RMS}} \rangle} = \frac{1}{1.71} \ln{\left( \frac{N}{0.590} \right)},
\end{equation}
and
\begin{equation}
{\langle \ln{D_{\rm RMS}} \rangle} = \frac{1}{1.68} \ln{\left( \frac{N}{0.275} \right)}.
\end{equation}
The fractal dimension reported in previous studies \cite{1989PhRvA..40..428T, meakin1999historical, 2009JPSJ...78a3605O} is $d_{\rm f} \approx 1.71$, which is consistent with our result.
To the best of our knowledge, the geodesic dimension $d_{\rm g} = 1.68$ has never been reported.
Although the difference between $d_{\rm f}$ and $d_{\rm g}$ is not significant, we expect that $d_{\rm g}$ is indeed smaller than $d_{\rm f}$ as in the case of three-dimensional DLA clusters (see below).

\begin{figure}
\centering
\includegraphics[width = \columnwidth]{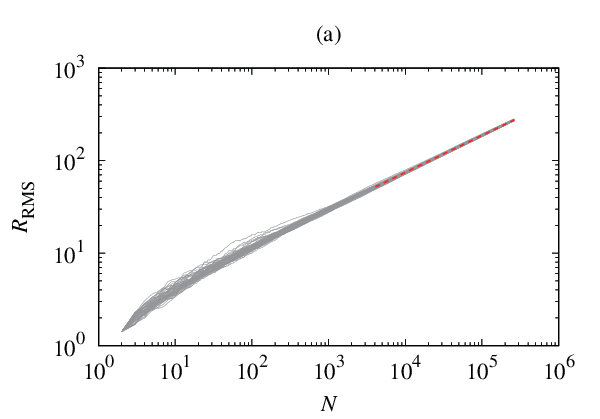}
\includegraphics[width = \columnwidth]{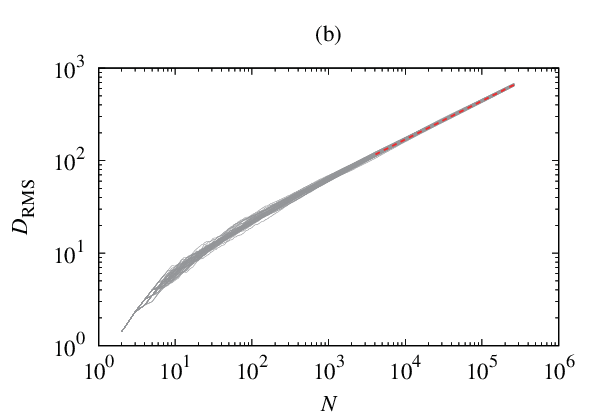}
\caption{\label{fig:N_R_D_DLA_3d}
Dependences of (a) $R_{\rm RMS}$ and (b) $D_{\rm RMS}$ on the number of constituent particles within aggregates formed by DLA in three-dimensional space.
Gray solid lines show the evolutionary paths of 50 aggregates, and the red dashed line represents the best fit obtained by the least-squares method.
(Color online)
}
\end{figure}

Figure \ref{fig:N_R_D_DLA_3d} shows the dependences of $R_{\rm RMS}$ and $D_{\rm RMS}$ on the number of constituent particles within aggregates formed by DLA in the three-dimensional space.
Gray solid lines show the evolutionary paths of 50 aggregates, and the red dashed line represents the best fit obtained by the least-squares method.
The best fits are ${\langle \ln{R_{\rm RMS}} \rangle} = 0.3996207(52) \ln{N} + 0.628283(60)$ and ${\langle \ln{D_{\rm RMS}} \rangle} = 0.4162007(61) \ln{N} + 1.298023(71)$, and the given uncertainties represent standard errors.
We obtain the following results:
\begin{equation}
{\langle \ln{R_{\rm RMS}} \rangle} = \frac{1}{2.50} \ln{\left( \frac{N}{0.208} \right)},
\end{equation}
and
\begin{equation}
{\langle \ln{D_{\rm RMS}} \rangle} = \frac{1}{2.40} \ln{\left( \frac{N}{0.0442} \right)}.
\end{equation}
The fractal dimension reported in previous studies \cite{meakin1999historical, 1989PhRvA..40..428T} is $d_{\rm f} \approx 2.50$, which is consistent with our result.
We note that $d_{\rm f} = 2.5$ is predicted from the mean-field theory \cite{1989PhRvA..40..428T}.
To the best of our knowledge, the geodesic dimension $d_{\rm g} = 2.40$ has never been reported; in this study, it is smaller than $d_{\rm f}$.

\section{Growing Self-Avoiding Walk}
\label{sec:GSAW}

\begin{figure}
\centering
\includegraphics[width = \columnwidth]{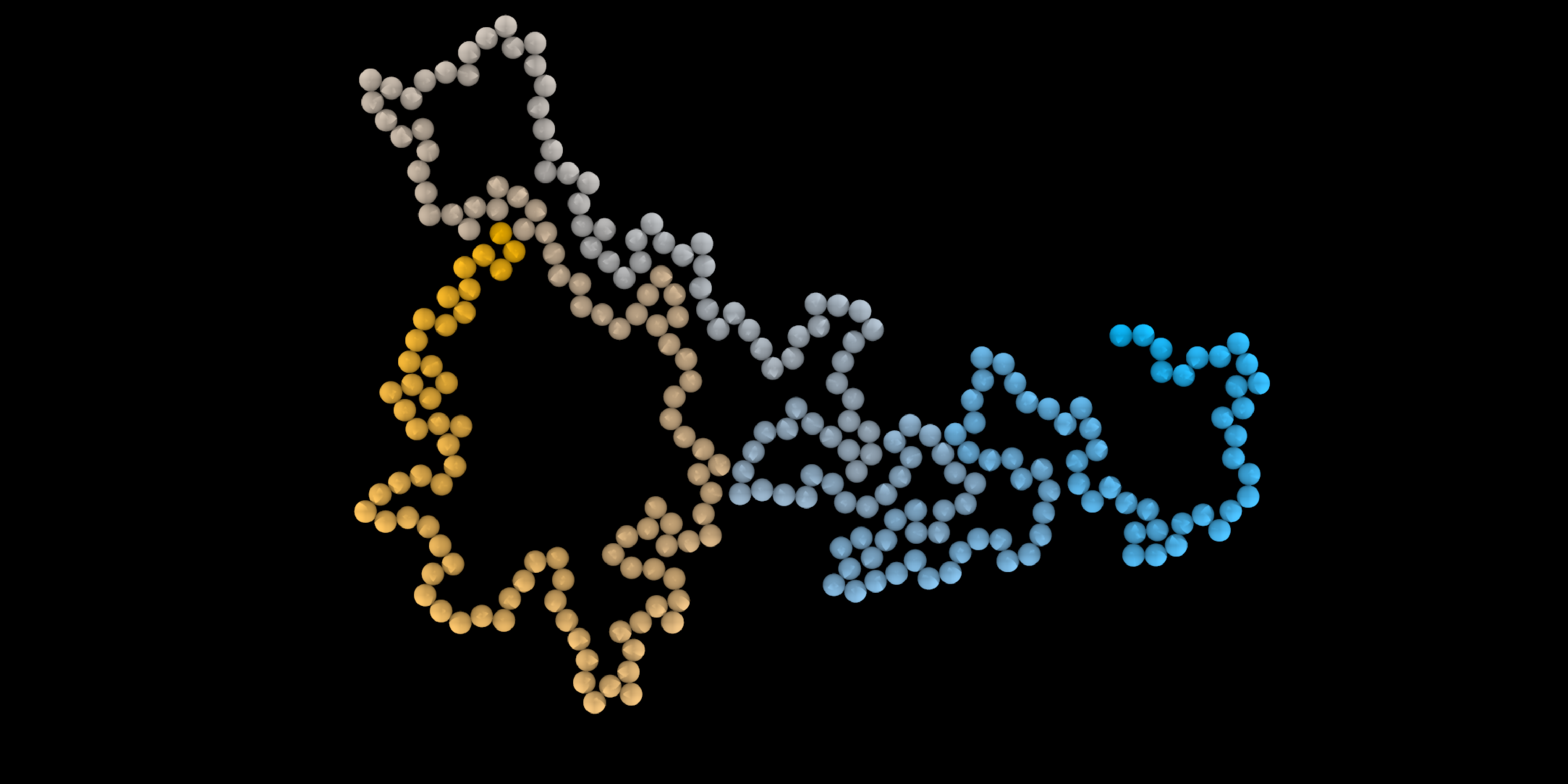}
\caption{\label{fig:SAW_2d}
Snapshot of a cluster formed by the GSAW process in two-dimensional space.
The color of each particle represents the timing of deposition; blue particles are deposited on the cluster at an earlier stage than yellow ones.
The growth of the cluster ends at the present state, and the number of constituent particles is $N = 265$.
(Color online)
}
\end{figure}

GSAW is a random walk sequence under the constraint that no two particles overlap each other \cite{1984PhRvL..52.1257M, 1986JPhA...19..279L, 2020PhRvE.102c2132H, 1984JChPh..81..584H}.
The GSAW process is widely used to model the scaling behavior of polymers \cite{1971JPhA....4L..82M}.
Note that a GSAW is terminated at a finite length when it reaches a state where there is no free space to add a new particle \cite{1984JChPh..81..584H}.
Figure \ref{fig:SAW_2d} shows a snapshot of a cluster formed by the GSAW process in the two-dimensional space.
In this sequence, the growth of the cluster ends at $N = 265$.

\begin{figure}
\centering
\includegraphics[width = \columnwidth]{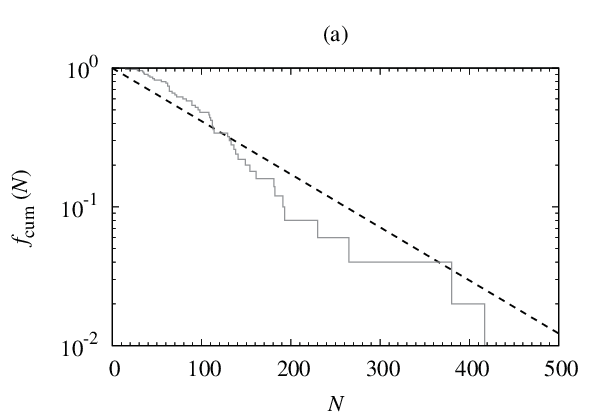}
\includegraphics[width = \columnwidth]{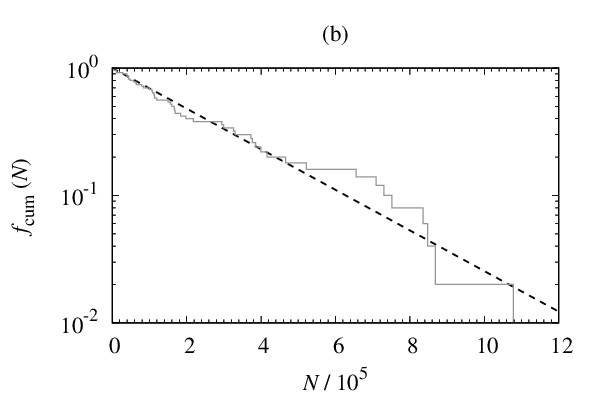}
\caption{\label{fig:N_frequency_SAW}
Complementary cumulative fraction of clusters that consist of $N$ particles or less at the termination state, $f_{\rm cum} {( N )}$.
(a) Two-dimensional GSAW clusters.
(b) Three-dimensional GSAW clusters.
We perform 50 runs of sequential growth of clusters for both two- and three-dimensional cases.
Dashed lines show the empirical fitting, $f_{\rm cum} {( N )} \sim \exp{\left( - N / N_{\rm end, mean} \right)}$.
}
\end{figure}

Here, we study the off-lattice GSAW process, although the GSAW process on the Cartesian lattice has mainly been investigated in previous studies \cite{1984PhRvL..52.1257M, 1986JPhA...19..279L, 1984JChPh..81..584H, 2020PhRvE.102c2132H}. 
First, we show the frequency distribution of the length of clusters at the termination state in Figure \ref{fig:N_frequency_SAW}.
Here, we define $f_{\rm cum} {( N )}$ as the complementary cumulative fraction of clusters that consist of $N$ or less particles at the termination state.
Figures \ref{fig:N_frequency_SAW}(a) and \ref{fig:N_frequency_SAW}(b) show results for two- and three-dimensional GSAW clusters, respectively.
In this study, we perform 50 runs of sequential growth of clusters for both two- and three-dimensional cases.

We define $N_{\rm end}$ as the number of constituent particles at the termination state.
The ensemble means of $N_{\rm end}$ are
\begin{equation}
N_{\rm end, mean} = 1.14 \times 10^{2}
\end{equation}
for the two-dimensional case and 
\begin{equation}
N_{\rm end, mean} = 2.73 \times 10^{5}
\end{equation}
for the three-dimensional case.
For the GSAW process on the Cartesian lattice, previous studies showed that $N_{\rm end, mean} \approx 72$ for two-dimensional clusters \cite{1984JChPh..81..584H, 2020PhRvE.102c2132H} and $N_{\rm end, mean} \approx 3.95 \times 10^{3}$ for three-dimensional clusters \cite{Pfoertner}.
The $N_{\rm end, mean}$ values we obtained for the off-lattice GSAW process are larger than those for the on-lattice GSAW process, reflecting the higher degree of freedom in the growth direction in the off-lattice case.
Indeed, the mean length of GSAWs on a square ladder (i.e., a lattice with infinite height and a width of 2) was exactly determined by Klotz and Sullivan \cite{2022arXiv220700539K} to be $N_{\rm end, mean} = 18$.

The differential frequency distribution of $N_{\rm end}$ has been reported for on-lattice GSAW \cite{Pfoertner, 2020PhRvE.102c2132H}, and it can be approximated by a simple exponential function except for the region where $N_{\rm end} \ll N_{\rm end, mean}$.
Then, we assume that the complementary cumulative fraction, $f_{\rm cum} {( N )}$, is approximately given as
\begin{equation}
f_{\rm cum} {( N )} \sim \exp{\left( - \frac{N}{N_{\rm end, mean}} \right)}.
\end{equation}
The dashed lines shown in Figure \ref{fig:N_frequency_SAW} roughly reproduce the trend of $f_{\rm cum} {( N )}$.
We can see this trend much clearer when the number of simulation runs is sufficiently large.

\begin{figure}
\centering
\includegraphics[width = \columnwidth]{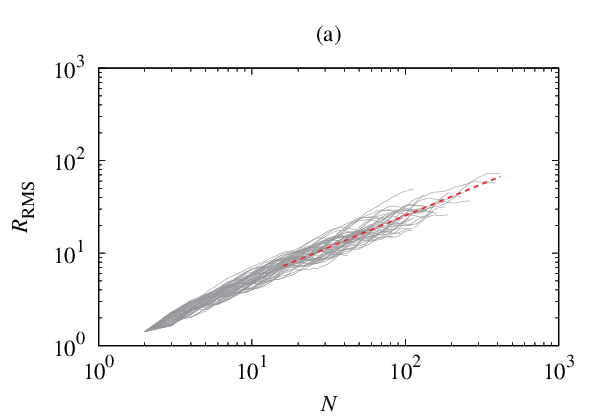}
\includegraphics[width = \columnwidth]{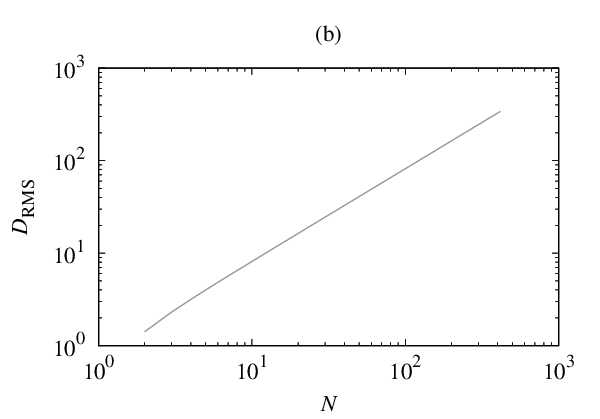}
\caption{\label{fig:N_R_D_SAW_2d}
Dependences of (a) $R_{\rm RMS}$ and (b) $D_{\rm RMS}$ on the number of constituent particles within aggregates formed by GSAW in two-dimensional space.
Gray solid lines show the evolutionary paths of 50 aggregates, and the red dashed line represents the best fit obtained by the least-squares method.
(Color online)
}
\end{figure}

We calculate the exponents ($d_{\rm f}$ and $d_{\rm g}$) and prefactors ($k_{\rm f}$ and $k_{\rm g}$) using the data for $N \ge 2^{4}$ for the two-dimensional case and $N \ge 2^{12}$ for the three-dimensional case.
Figure \ref{fig:N_R_D_SAW_2d} shows the dependences of $R_{\rm RMS}$ and $D_{\rm RMS}$ on the number of constituent particles within aggregates formed by GSAW in the two-dimensional space.
Gray solid lines show the evolutionary paths of 50 aggregates, and the red dashed line represents the best fit obtained by the least-squares method: ${\langle \ln{R_{\rm RMS}} \rangle} = 0.6833(36) \ln{N} + 0.088(16)$, where the given uncertainties represent standard errors.
We obtain the following result for ${\langle \ln{R_{\rm RMS}} \rangle}$:
\begin{equation}
{\langle \ln{R_{\rm RMS}} \rangle} = \frac{1}{1.46} \ln{\left( \frac{N}{0.879} \right)}.
\end{equation}
For the two-dimensional GSAW clusters on the Cartesian lattice, $1 / d_{\rm f} \approx 0.68$ was reported in previous studies \cite{1986JPhA...19..279L, 2020PhRvE.102c2132H}.
As $1 / 1.46 \approx 0.68$, the fractal dimension of off-lattice GSAW clusters obtained from our numerical simulations would be equal to that for on-lattice clusters.

The exact solution of $D_{\rm RMS}$ for GSAW clusters is
\begin{equation}
D_{\rm RMS} = \sqrt{\frac{2 {\left( N^{2} - 1 \right)}}{3}}.
\end{equation}
For a sufficiently large $N$, it is approximately given by
\begin{equation}
\ln{D_{\rm RMS}} \approx \ln{\left( \frac{N}{\sqrt{3/2}} \right)}.
\end{equation}

\begin{figure}
\centering
\includegraphics[width = \columnwidth]{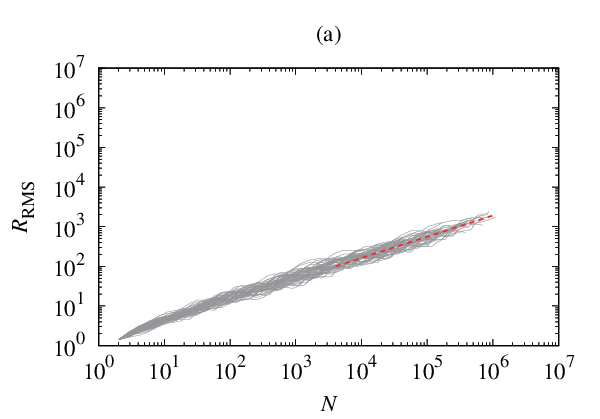}
\includegraphics[width = \columnwidth]{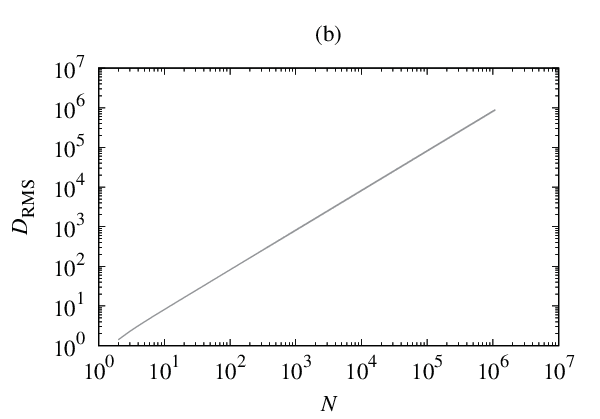}
\caption{\label{fig:N_R_D_SAW_3d}
Dependences of (a) $R_{\rm RMS}$ and (b) $D_{\rm RMS}$ on the number of constituent particles within aggregates formed by GSAW in three-dimensional space.
Gray solid lines show the evolutionary paths of 50 aggregates, and the red dashed line represents the best fit obtained by the least-squares method.
(Color online)
}
\end{figure}

Figure \ref{fig:N_R_D_SAW_3d} shows the dependences of $R_{\rm RMS}$ and $D_{\rm RMS}$ on the number of constituent particles within aggregates formed by GSAW in the three-dimensional space.
Gray solid lines show the evolutionary paths of 50 aggregates, and the red dashed line represents the best fit obtained by the least-squares method: ${\langle \ln{R_{\rm RMS}} \rangle} = 0.534744(52) \ln{N} + 0.16841(63)$, where the given uncertainties represent standard errors.
We obtain the following result for ${\langle \ln{R_{\rm RMS}} \rangle}$:
\begin{equation}
{\langle \ln{R_{\rm RMS}} \rangle} = \frac{1}{1.87} \ln{\left( \frac{N}{0.730} \right)}.
\end{equation}

\section{Vicsek Fractals}
\label{sec:Vicsek}

\begin{figure}
\centering
\includegraphics[width = \columnwidth]{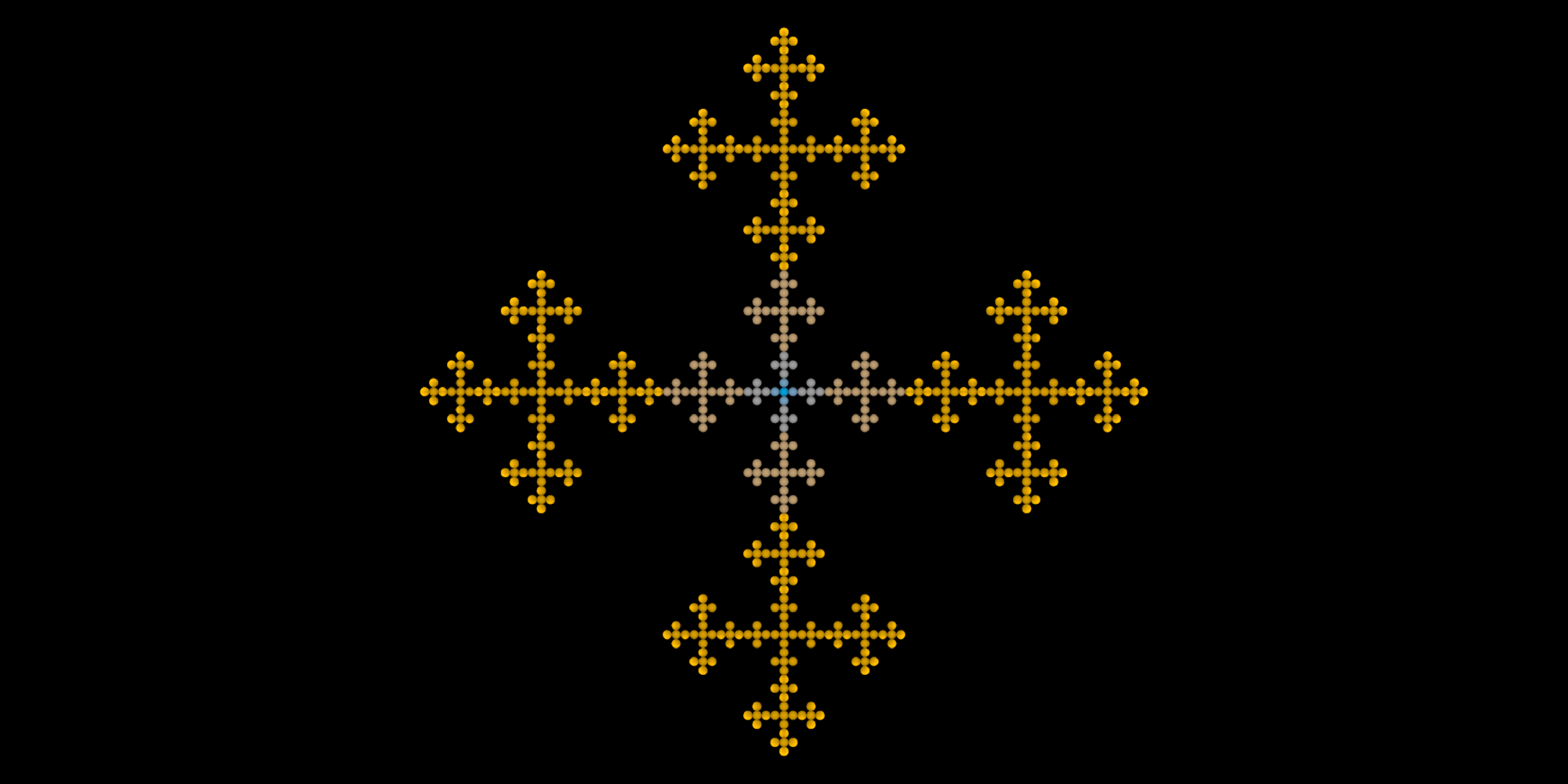}
\caption{\label{fig:Vicsek_2d}
Snapshot of a two-dimensional Vicsek fractal cluster for $N = 625$.
The color of each particle represents the timing of deposition; blue particles are deposited on the cluster at an earlier stage than yellow ones.
(Color online)
}
\end{figure}

The concept of fractals plays a pivotal role in characterizing the complex nature of aggregation.
Although the random clusters in nature are formed by probabilistic processes, structural and dynamical properties of regular fractals formed by deterministic processes have also been studied \cite{zbMATH02653553, 2003PhRvE..67f1103B, 2008JPhA...41V5102Z, 2015PhRvE..92f2146B, 2020PhyS...95f5210D}.
The Vicsek fractal cluster \cite{1983JPhA...16L.647V} is one of the most common regular fractal clusters.
Figure \ref{fig:Vicsek_2d} is a snapshot of a Vicsek fractal cluster in the two-dimensional space.

A Vicsek fractal cluster is formed by adding its clone clusters to the edge of the original cluster.
In this study, we consider conventional Vicsek fractals, i.e., Vicsek fractals on the Cartesian lattice.
The numbers of constituent particles at the $i$th step are $N = 5^{i}$ for the cluster in the two-dimensional space and $N = 7^{i}$ in the three-dimensional space.

\begin{figure}
\centering
\includegraphics[width = \columnwidth]{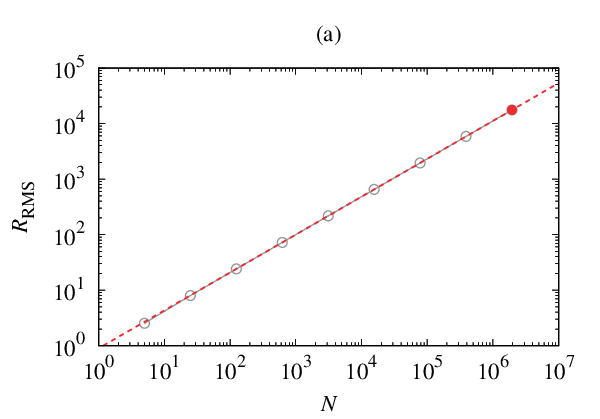}
\includegraphics[width = \columnwidth]{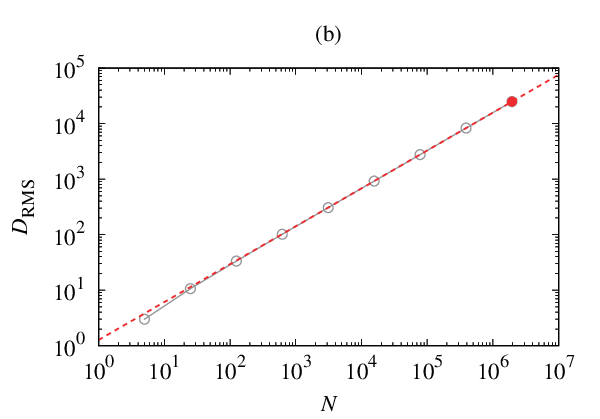}
\caption{\label{fig:N_R_D_Vicsek_2d}
Dependences of (a) $R_{\rm RMS}$ and (b) $D_{\rm RMS}$ on the number of constituent particles within two-dimensional Vicsek fractal aggregates.
The gray solid line shows the evolutionary path, and the red dashed line represents the fit by the value at $N = 5^{9}$.
(Color online)
}
\end{figure}

Figure \ref{fig:N_R_D_Vicsek_2d} shows the dependences of $R_{\rm RMS}$ and $D_{\rm RMS}$ on the number of constituent particles within Vicsek aggregates in the two-dimensional space.
The gray solid line shows the evolutionary path, and the red dashed line represents the fit by the value at $N = 5^{9}$.
We found that both the fractal and geodesic dimensions are $d_{\rm f} = d_{\rm g} = \log_{3}{5} \approx 1.46$.
Using the values at $N = 5^{9}$, we obtain the following results:
\begin{equation}
R_{\rm RMS} = {\left( \frac{N}{1.18} \right)}^{1 / \log_{3}{5}},
\end{equation}
and
\begin{equation}
D_{\rm RMS} = {\left( \frac{N}{0.709} \right)}^{1 / \log_{3}{5}}.
\end{equation}
Both the fractal and geodesic dimensions are $d_{\rm f} = d_{\rm g} = \log_{3}{5} \approx 1.46$ for Vicsek aggregates in the two-dimensional space.

\begin{figure}
\centering
\includegraphics[width = \columnwidth]{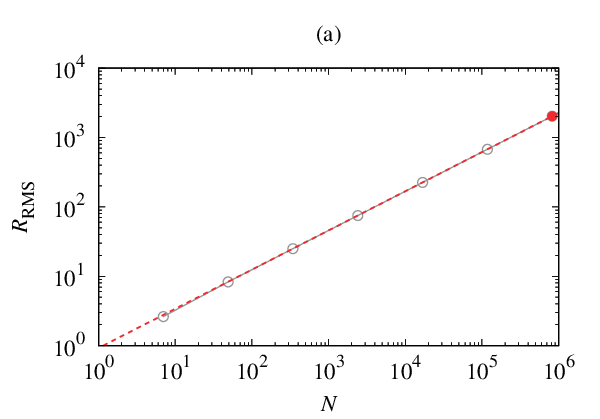}
\includegraphics[width = \columnwidth]{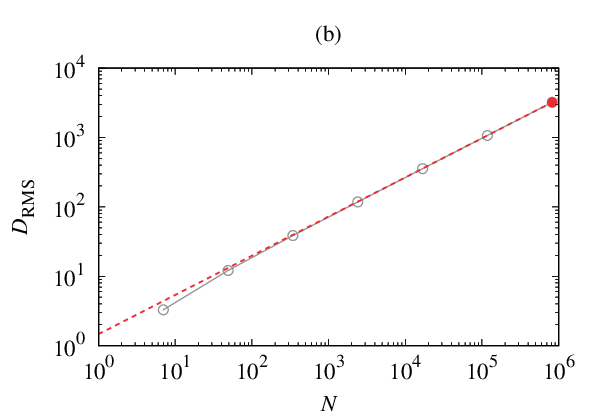}
\caption{\label{fig:N_R_D_Vicsek_3d}
Dependences of (a) $R_{\rm RMS}$ and (b) $D_{\rm RMS}$ on the number of constituent particles within three-dimensional Vicsek fractal aggregates.
The gray solid line shows the evolutionary path, and the red dashed line represents the fit by the value at $N = 7^{7}$.
(Color online)
}
\end{figure}

Figure \ref{fig:N_R_D_Vicsek_3d} shows the dependences of $R_{\rm RMS}$ and $D_{\rm RMS}$ on the number of constituent particles within Vicsek aggregates in the three-dimensional space.
Using the values at $N = 5^{9}$, we obtain the following results:
\begin{equation}
R_{\rm RMS} = {\left( \frac{N}{1.15} \right)}^{1 / \log_{3}{7}},
\end{equation}
and
\begin{equation}
D_{\rm RMS} = {\left( \frac{N}{0.507} \right)}^{1 / \log_{3}{7}}.
\end{equation}
Both the fractal and geodesic dimensions are $d_{\rm f} = d_{\rm g} = \log_{3}{7} \approx 1.77$ for Vicsek aggregates in the three-dimensional space.

For Vicsek fractals, we found that $d_{\rm f} = d_{\rm g}$ for both two- and three-dimensional cases.
In contrast, $d_{\rm f} > d_{\rm g}$ for the other three types of fractal investigated in this study.
The difference between fractal and geodesic dimensions varies significantly depending on the preparation procedure for fractals.

Note that $d_{\rm f}$ is not equal to $d_{\rm g}$ for some deterministic clusters.
For example, a two-dimensional spiral pattern with constant separation has the fractal and geodesic dimensions of $d_{\rm f} = 2$ and $d_{\rm g} = 1$, respectively.
This finding might indicate that the differences between $d_{\rm f}$ and $d_{\rm g}$ found for BCCA, DLA, and GSAW clusters neither originate from their probabilistic nature nor represent the degree of freedom for coagulation processes.
The branching pattern might be the key to understanding the $d_{\rm g}$ of fractals, and we will investigate it in detail in the future.

\section{Summary}
\label{sec:summary}

In this study, we investigated the fractal and geodesic dimensions $d_{\rm f}$ and $d_{\rm g}$ for aggregates formed by various preparation procedures.
We determined $d_{\rm f}$ and $d_{\rm g}$ for BCCA, DLA, and GSAW clusters in two- and three-dimensional spaces, and we also performed calculations for Vicsek fractals on the Cartesian lattice.

\begin{figure}
\centering
\includegraphics[width = \columnwidth]{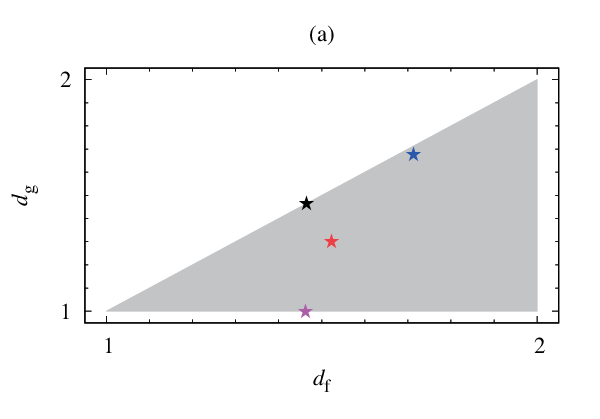}
\includegraphics[width = \columnwidth]{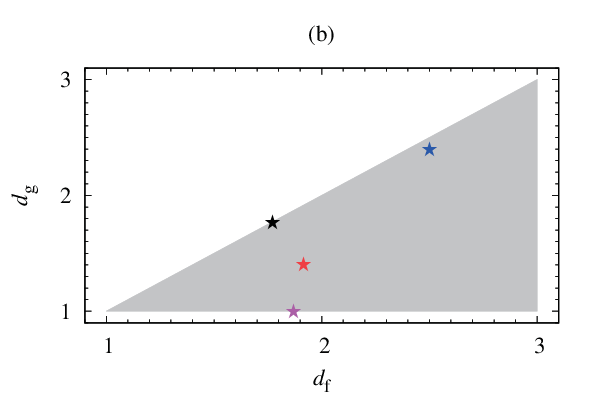}
\caption{\label{fig:df_dg}
Fractal and geodesic dimensions of fractal aggregates, $d_{\rm f}$ and $d_{\rm g}$, formed by BCCA (red), DLA (blue), GSAW (magenta), and the Vicsek model (black).
(a) Two-dimensional clusters.
(b) Three-dimensional clusters.
The gray shaded region represents the parameter range of two exponents: $1 \le d_{\rm g} \le d_{\rm f} \le d_{\rm space}$, where $d_{\rm space} = 2$ or $3$ is the spatial dimension. 
(Color online)
}
\end{figure}

We plot our results on the $d_{\rm f}$--$d_{\rm g}$ plane in Figure \ref{fig:df_dg}.
The exponents satisfy $1 \le d_{\rm g} \le d_{\rm f} \le d_{\rm space}$, where $d_{\rm space} = 2$ or $3$ is the spatial dimension.
Our results show that the pairs of $( d_{\rm f}, d_{\rm g} )$ are not clustered in a narrow area but are spread out across a wide space on the $d_{\rm f}$--$d_{\rm g}$ plane.
The exponents depend on the preparation procedure for aggregates; however, their relationship remains not well clarified.
Future studies on this topic are of great importance.

Although we have not quantified the dispersion of $\ln{R_{\rm RMS}}$, our findings indicate a smaller dispersion for DLA clusters than for BCCA and GSAW clusters.
The variability in $\ln{R_{\rm RMS}}$ (and $\ln{D_{\rm RMS}}$) is likely associated with variations in their thermal and mechanical properties.
Consequently, a quantitative analysis of the dispersion will be imperative for future studies.

\begin{acknowledgments}

We thank Mikito Furuichi for the fruitful discussion.
We also thank the anonymous reviewer for providing a constructive review to improve this paper.
Numerical computations were carried out on PC cluster at CfCA, NAOJ.

\end{acknowledgments}


\bibliography{apssamp}
\bibliographystyle{jpsj}

\end{document}